\documentclass[10pt,epsfig]{article}  
\usepackage{epsfig}
\usepackage[body={6.5in, 9.0in},left=1.0in, top = 1.0in]{geometry}
\usepackage{amsmath}
\usepackage{amssymb}
\usepackage{graphicx}
\begin{document}
\newcommand{\sheptitle}
{Dependence of $\tan^2 \theta_{12}$ on Dirac CP phase $\delta$ in tri-bimaximal
 neutrino mixing under charged lepton correction}
\newcommand{\shepauthor}
{Chandan Duarah$^a$, Abhijeet Das$^b$ and N. Nimai Singh$^{c}$\footnote{\it{Corresponding author:} nimai03@yahoo.com}}
\newcommand{\shepaddress}
   { $^{a,c}$Department of Physics, Gauhati University,
     Guwahati-781014, India \\
   $^b$Department of Physics, Assam University, Diphu Campus, Diphu-782460, Assam, India}
\newcommand{\shepabstract}
{We consider charged lepton correction to Tri-bimaximal(TBM) neutrino mixing,
defined by the relation $U_{PMNS}=U_l^{\dagger} U_{TB}$ and find possible form
of $U_l$ which can impart non-zero value of $\sin \theta_{13}$ as well as
$\tan^2 \theta_{23}<1$, consistent with latest global analysis data. We adopt 
a new parametrization, other than the standard PDG parametrization, to introduce 
 Dirac CP violating phase $\delta$ in the PMNS matrix which is discussed by
 Fritzsch. Under such charged lepton correction pattern we note that $\tan^2 \theta_{12}$
becomes dependent
on the CP phase $\delta$ from where constraints on $\delta$ phase can be
obtained after employing experimental range of mixing angles.
To compute the values of mixing angles we assume the charged lepton
correction to be of Cabibbo-Kobayashi-Maskawa(CKM) like. Since all the
mixing matrices, involved in the calculation, are derived from three dimensional
rotation matrices they satisfy unitarity condition. \\

Key-words: Charged lepton correction, Tri-bimaximal mixing.\\
PACS number: 14.60 Pq}
\begin{titlepage}
\begin{flushright}
\end{flushright}
\begin{center}
{\large{\bf\sheptitle}}
\bigskip\\
\shepauthor
\\
{\it\shepaddress}\\
\vspace{.5in}
{\bf Abstract}
\bigskip
\end{center}
\setcounter{page}{0}
\shepabstract
\end{titlepage}

\section{Introduction}
Recent precision measurements[1-4] and latest global $3\nu$ oscillation analysis
of neutrino mixing parameters by two individual groups[5,6], have confirmed 
the non-vanishing value of
$\theta_{13}$, and also predict a bestfit value of $\sin \theta_{13}$ which
lies near $\displaystyle \frac{\lambda}{\sqrt{2}}$, $\lambda$ being the 
Wolfenstein parameter. Further, both groups of the global analysis provide
an indication for $\theta_{23}$ to lie in the first octant($\displaystyle
\theta_{23}<\frac{\pi}{4}$) for normal hierarchy (NH) upto $1\sigma$ range
of data. One of the important aspects of netrino physics is to understand such
mixing patterns[7]. Tri-bimaximal(TBM)[8] neutrino mixing is the most popular mixing
pattern of neutrinos among several special mixings obeying $\mu-\tau$
symmetry, whose predictions are attractively close to global data. It is therefore likely
to believe that recent global data could have been accomodated within TBM mixing
under certain perturbation to itself. Charged lepton corrections[9,10,11] in this context,   
is an attractive tool which can generate desired results.\\

The lepton mixing matrix known as 
Pontecorvo-Maki-Nakagawa-Sakata(PMNS) matrix[12], is usually expressed as 
\begin{equation}
U_{PMNS}=U_l^{\dagger} U_{\nu},
\end{equation} 
where $U_{l}$ and $U_{\nu}$ are the diagonalizing matrices 
for charged lepton and left-handed Majorana neutrino mass matrices
respectively. They are defined through the relations :
$m_l=U_{lL} m_l^{diag} V_{lR}^{\dagger}$ and $m_{\nu}=U_{\nu}^{*}
m_{\nu}^{diag} U_{\nu}^{\dagger}$, where $m_l^{diag}=Diag(m_e,m_{\mu},m_{\tau})$
and $m_{\nu}^{diag}=Diag(m_1,m_2,m_3)$. In the basis where charged lepton mass matrix
$m_l$ is diagonal, $U_{PMNS}=U_{\nu}$, $U_l$ being identity matrix, and
the left-handed Majorana mass matrix is then expressible as[15], 
$m_{\nu}^{\prime}=U_{lL}^{\dagger} m_{\nu} U_{lL}$. The PMNS matrix is also
analogous to the CKM matrix, 
$V_{CKM}=U_{uL}^{\dagger} U_{dL}$ for quark sector[13,14], where $U_{uL}$
and $U_{dL}$ are the diagonalizing matrices for up-type and down-type
quark mass matrices.\\

In the standard Particle Data Group (PDG) parametrization[14],
PMNS matrix can be parametrized as 
\begin{equation}
U_{PMNS}=R_{23}.U_{13}.R_{12}.P,
\end{equation}
where,
\begin{equation}
R_{12}= \begin{pmatrix}
            c_{12} & s_{12} & 0 \\
           -s_{12} & c_{12} & 0 \\
              0    &    0   & 1 \\

                \end{pmatrix}, \ \ R_{23}= \begin{pmatrix}
                                         1      & 0      & 0 \\
                                        0       & c_{23} & s_{23} \\
                                           0    & -s_{23} & c_{23} \\

                                             \end{pmatrix},
\end{equation}
\begin{equation}
U_{13}= \begin{pmatrix}
            c_{13} & 0     & s_{13} e^{-i \delta} \\
               0    & 1     & 0 \\
     -s_{13} e^{i \delta}  &    0   & c_{13} \\

                \end{pmatrix}.
\end{equation}
and $P=diag(1, e^{i \alpha}, e^{i \beta})$. Here $c_{ij}=\cos \theta_{ij}$ and
$s_{ij}=\sin \theta_{ij}$ with
$\theta_{12}$, $\theta_{23}$ and
$\theta_{13}$ being the solar angle, atmospheric angle and the reactor angle
respectively. $\delta$ is the Dirac CP violating phase while $\alpha$ and
$\beta$ are the two Majorana CP violating phases. Then eq.(2) yields the
following standard form of the PMNS matrix : 
\begin{equation}
       U_{PMNS} = \begin{pmatrix}
    c_{12} c_{13}                       & s_{12} c_{13} 
                                                          & s_{13} e^{-i \delta}\\
    -s_{12} c_{23}-c_{12} s_{23} s_{13}e^{i \delta} & c_{12} c_{23}-s_{12} s_{23} s_{13} e^{i \delta}
                                                          & s_{23} c_{13}\\
    s_{12} s_{23}-c_{12} c_{23} s_{13}e^{i \delta} & -c_{12} s_{23}-s_{12} c_{23} s_{13} e^{i \delta} 
                                                          & c_{23} c_{13} \\ 
                                                      \end{pmatrix}.P.
\end{equation}
We would now like to drop the Majorana phase matrix $P$ in our discussion.
Then from the PMNS matrix in eq.(5) we obtain the
following useful expressions for mixing angles:
\begin{equation}
\sin^2 \theta_{13}=|U_{e3}|^2,
\end{equation}
\begin{equation}
\tan^2 \theta_{12}=\displaystyle \frac{|U_{e2}|^2}{|U_{e1}|^2},
\end{equation}
\begin{equation}
\tan^2 \theta_{23}=\displaystyle \frac{|U_{\mu 2}|^2}{|U_{\tau 3}|^2},
\end{equation}
which are free from the Dirac CP violating phase $\delta$. \\

The TBM mixing matrix is now followed from eq.(2) with $\displaystyle s_{12}=
\frac{1}{\sqrt{3}}$, $\displaystyle s_{23}=\frac{1}{\sqrt{2}}$ and $ s_{13}=0$
and is given by
\begin{equation}
U_{TB}= \begin{pmatrix}
          \sqrt{\frac{2}{3}} &  \frac{1}{\sqrt{3}} & 0 \\
           -\frac{1}{\sqrt{6}}  & \frac{1}{\sqrt{3}}  & \frac{1}{\sqrt{2}} \\
          \frac{1}{\sqrt{6}}  & -\frac{1}{\sqrt{3}}   & \frac{1}{\sqrt{2}} \\

                \end{pmatrix}.
\end{equation}
The CP phase $\delta$ disappears along with $s_{13}$ in eq.(9). However it can be 
restored by adopting another parametrization[16] where $U_{13}$ in eq.(4) is replaced
by
\begin{equation}
U^{\prime}_{13}= \begin{pmatrix}
            c_{13} e^{i \delta} & 0     & s_{13}  \\
               0    & 1     & 0 \\
          -s_{13}   &    0   & c_{13} e^{-i \delta} \\

                \end{pmatrix}.
\end{equation}
Under this parametrization, the new TBM matrix becomes
\begin{equation}
U^{\prime}_{TB}= \begin{pmatrix}
          \sqrt{\frac{2}{3}} e^{i \delta} &  \frac{1}{\sqrt{3}} e^{i \delta} & 0 \\
           -\frac{1}{\sqrt{6}}  & \frac{1}{\sqrt{3}}  & \frac{1}{\sqrt{2}} e^{-i \delta} \\
          \frac{1}{\sqrt{6}}  & -\frac{1}{\sqrt{3}}   & \frac{1}{\sqrt{2}} e^{-i \delta} \\

                \end{pmatrix},
\end{equation}
and the general PMNS matrix looks like
\begin{equation}
       U^{\prime}_{PMNS} = \begin{pmatrix}
    c_{12} c_{13} e^{i \delta}  & s_{12} c_{13} e^{i \delta}
                                               & s_{13} \\
    -s_{12} c_{23}-c_{12} s_{23} s_{13} & c_{12} c_{23}-s_{12} s_{23} s_{13} 
                                               & s_{23} c_{13}e^{-i \delta}\\
    s_{12} s_{23}-c_{12} c_{23} s_{13} & -c_{12} s_{23}-s_{12} c_{23} s_{13}  
                                               & c_{23} c_{13}e^{-i \delta} \\ 
                                                      \end{pmatrix}.
\end{equation}
The phase $\delta$ in eqs. (11) and (12) has no physical significance. We
mention them only in the context of the parametrization (10) as we would like to 
adopt this parametrization in our future calculations.\\

The paper is organized as follows: Section 2 is divided into three subsections.
In subsection 2.1 we begin the discussion of charged lepton correction to
TBM mixing in the absence of CP violation. In subsection 2.2 we introduce
the Dirac CP violating phase into the discussion and present the central
issue of the present paper. Finally section 3 is devoted to summary 
and discussion. 

\section{Charged lepton correction to TBM mixing}
\subsection{Without Dirac CP phase}

We begin with eq.(1) where $U_{\nu}$ is to be given by $U_{TB}$ in eq.(9)
for our case. We then consider the following form of the charged lepton mixing matrix:
\begin{equation}
U_{l}=\tilde{R}_{23}.\tilde{R}_{12},
\end{equation}
with
\begin{equation}
\tilde{R}_{12}= \begin{pmatrix}
            \tilde{c}_{12} & \tilde{s}_{12} & 0 \\
           -\tilde{s}_{12} & \tilde{c}_{12} & 0 \\
              0    &    0   & 1 \\

                \end{pmatrix}, \ \ \tilde{R}_{23}= \begin{pmatrix}
                                         1      & 0      & 0 \\
                                        0   & \tilde{c}_{23} & \tilde{s}_{23} \\
                                       0    & -\tilde{s}_{23} & \tilde{c}_{23} \\

                                             \end{pmatrix}.
\end{equation}
The structure of $U_l$ defined by eq.(13) is analogous to the that of $U_{TB}$ in eq.(9)
in the sense that $U_{TB}$ is also given by $U_{TB}=R_{23}.R_{12}$ with $\displaystyle
s_{12}=\frac{1}{\sqrt{3}}$ and $\displaystyle s_{23}=\frac{1}{\sqrt{2}}$. 
Then eqs. (13) and (14) yield
\begin{equation}
U_l= \begin{pmatrix}
         \tilde{c}_{12} & \tilde{s}_{12}   & 0 \\
     -\tilde{s}_{12}\tilde{c}_{23} & \tilde{c}_{12} \tilde{c}_{23} &  \tilde{s}_{23} \\
     \tilde{s}_{12}\tilde{s}_{23} & -\tilde{c}_{12}\tilde{s}_{23}   & \tilde{c}_{23}  \\

                \end{pmatrix}.
\end{equation}
With this $U_l$ we get the PMNS matrix, from the relation $U_{PMNS}=U^{\dagger}_l U_{TB}$,
as
\begin{equation}
U_{PMNS}= \begin{pmatrix}
       \sqrt{\frac{2}{3}}[\tilde{c}_{12}+ \frac{1}{2}\tilde{s}_{12}
                                 (\tilde{c}_{23}+\tilde{s}_{23})]  
             & \frac{1}{\sqrt{3}}[\tilde{c}_{12}-\tilde{s}_{12}(\tilde{c}_{23}+\tilde{s}_{23})]  
                & -\frac{1}{\sqrt{2}}\tilde{s}_{12}(\tilde{c}_{23}-\tilde{s}_{23}) \\
        -\frac{1}{\sqrt{6}}[\tilde{c}_{12}(\tilde{c}_{23}+\tilde{s}_{23})-2\tilde{s}_{12}] 
           & \frac{1}{\sqrt{3}}[\tilde{s}_{12}+\tilde{c}_{12}(\tilde{c}_{23}+\tilde{s}_{23})]  
                 & \frac{1}{\sqrt{2}}\tilde{c}_{12}(\tilde{c}_{23}-\tilde{s}_{23}) \\
              \frac{1}{\sqrt{6}}(\tilde{c}_{23}-\tilde{s}_{23})    
              &    -\frac{1}{\sqrt{3}}(\tilde{c}_{23}-\tilde{s}_{23})  
                  & \frac{1}{\sqrt{2}}(\tilde{c}_{23}+\tilde{s}_{23}) \\

                \end{pmatrix}.
\end{equation}

\begin{table}[tbp]
\begin{center}
\begin{tabular}{|c|c|c|c|}
\hline
parameter & best fit  &    $1\sigma$ range    & $3\sigma$ range \\  
     & Ref[5] \ \ \ Ref[6] & Ref[5] \ \ \ \ \ \ \ Ref[6]
                                      & Ref[5] \ \ \ \ \ \ \ Ref[6] \\ \hline                                                 
$\tan^2 \theta_{12}$ & 0.443 \ \ \ 0.470 & 0.410-0.481 \ \ \ 0.435-0.506 & 0.350-0.560 \ \ \
                                                                   0.370-0.587  \\
$\tan^2 \theta_{23}$ & 0.629 \ \ \ 0.745 & 0.575-0.695 \ \ \ 0.667-0.855 & 0.495-1.755 \ \ \
                                                                         0.563-2.125 \\
$\sin^2 \theta_{13}$ & 0.0241 \ \ \ 0.0246 & 0.0216-0.0266 \ \ \ 0.0218-0.0275 
                                     & 0.0169-0.0313 \ \ \ 0.017-0.033 \\
 \hline
\end{tabular}
\caption{Best fit, $1\sigma$ and $3\sigma$ ranges of parameters for NH obtained
         from global analysis by Fogli {\it et al.}[5] and Forero {\it et al.}[6]}
\end{center}
\end{table}

This PMNS matrix predicts 
\begin{equation}
\sin^2 \theta_{13}=\displaystyle \frac{\tilde{s}^2_{12}(\tilde{c}_{23}-\tilde{s}_{23})^2}{2},
\end{equation}
\begin{equation}
\tan^2 \theta_{12}=\displaystyle \frac{1}{2} 
\left[\frac{\tilde{c}_{12}-\tilde{s}_{12}(\tilde{c}_{23}+\tilde{s}_{23})}
{\tilde{c}_{12}+\frac{1}{2}\tilde{s}_{12}(\tilde{c}_{23}+\tilde{s}_{23})}\right]^2,
\end{equation}
\begin{equation}
\tan^2 \theta_{23}=\displaystyle \frac{\tilde{c}^2_{12}(\tilde{c}_{23}-\tilde{s}_{23})^2}
{(\tilde{c}_{23}+\tilde{s}_{23})^2}.
\end{equation}

\begin{figure}
\begin{center}
\includegraphics[scale=1]{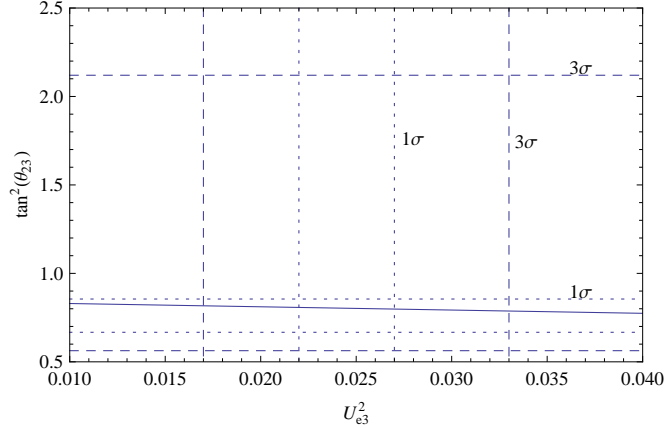}
\caption{Variation of $\tan^2 \theta_{23}$ with $U_{e3}^2$ for TBM
  mixing under charged lepton correction. Dotted and dashed lines
  represents $1\sigma$ and $3\sigma$ bounds respectively, 
  obtained from the global analysis[6]}
\end{center}
\end{figure}

To compute the numerical predictions let us now assume that the charged 
lepton corrections are Cabibbo-Kobayashi-Maskawa(CKM) like[14], which allows 
us to take
\begin{equation}
\tilde{s}_{12}= \lambda \ \  and \ \ \tilde{s}_{23}=A \lambda^2,
\end{equation}
where $\lambda$ is the Wolfestein parameter and is related to the Cabibbo angle 
($\theta_C$) by $\lambda=\sin \theta_C$. $A$ is a constant. Taking 
$\tilde{s}_{23} \approx 0.041$ with $\lambda=0.232$
and $A=0.759$ we get $\sin^2 \theta_{13}=0.0247$, $\tan^2 \theta_{12}=0.224$ and
$\tan^2 \theta_{23}=0.80$. The prediction on solar angle $\tan^2 \theta_{12}$ is 
significantly smaller than the global best fit value(Table-1). This is the problem
with TBM mixing under the charged lepton correction pattern considered.

\subsection{Dirac CP violation}
We now introduce the Dirac type CP violating phase $\delta$ in the PMNS matrix 
(16) by adopting the parametrization described in eq.(10).
Then the new PMNS matrix with Dirac CP phase is given by 
\begin{equation}
U^{\prime}_{PMNS}=U^{\dagger}_l(R_{23}U^{\prime}_{13}R_{12})=U^{\dagger}_l U^{\prime}_{TB}, 
\end{equation}
where $U_l$ and $U^{\prime}_{13}$ are respectively given by eqs. (15) and (10)
and we set $\displaystyle s_{12}=\frac{1}{\sqrt{3}}$, $\displaystyle s_{23}=
\frac{1}{\sqrt{2}}$ and $ s_{13}=0$. Thus we obtain
\begin{equation}
U^{\prime}_{PMNS}= \begin{pmatrix}
       \sqrt{\frac{2}{3}}[\tilde{c}_{12}e^{i \delta}+ \frac{1}{2}\tilde{s}_{12}
                                 (\tilde{c}_{23}+\tilde{s}_{23})]  
     & \frac{1}{\sqrt{3}}[\tilde{c}_{12}e^{i \delta}-\tilde{s}_{12}(\tilde{c}_{23}+\tilde{s}_{23})]  
           & -\frac{1}{\sqrt{2}}\tilde{s}_{12}(\tilde{c}_{23}-\tilde{s}_{23})e^{-i \delta} \\
   -\frac{1}{\sqrt{6}}[\tilde{c}_{12}(\tilde{c}_{23}+\tilde{s}_{23})-2\tilde{s}_{12}e^{i \delta}] 
  & \frac{1}{\sqrt{3}}[\tilde{s}_{12}e^{i \delta}+\tilde{c}_{12}(\tilde{c}_{23}+\tilde{s}_{23})]  
            & \frac{1}{\sqrt{2}}\tilde{c}_{12}(\tilde{c}_{23}-\tilde{s}_{23})e^{-i \delta} \\
              \frac{1}{\sqrt{6}}(\tilde{c}_{23}-\tilde{s}_{23})    
              &    -\frac{1}{\sqrt{3}}(\tilde{c}_{23}-\tilde{s}_{23})  
                  & \frac{1}{\sqrt{2}}(\tilde{c}_{23}+\tilde{s}_{23})e^{-i \delta} \\

                \end{pmatrix}.
\end{equation}

\begin{figure}
\begin{center}
\includegraphics[scale=1]{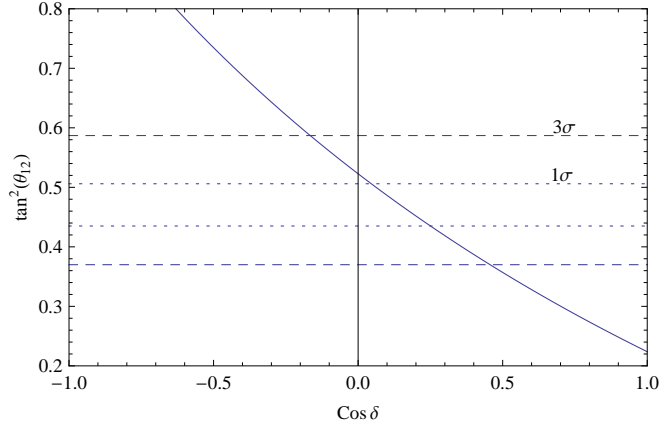}
\caption{Variation of $\tan^2 \theta_{12}$ with $\cos \delta$ for TBM
  mixing under charged lepton correction. Horizontal dotted and dashed lines
  represents $1\sigma$ and $3\sigma$ bounds respectively, 
  obtained from the global analysis[6]}
\end{center}
\end{figure}

\begin{figure}
\begin{center}
\includegraphics[scale=1]{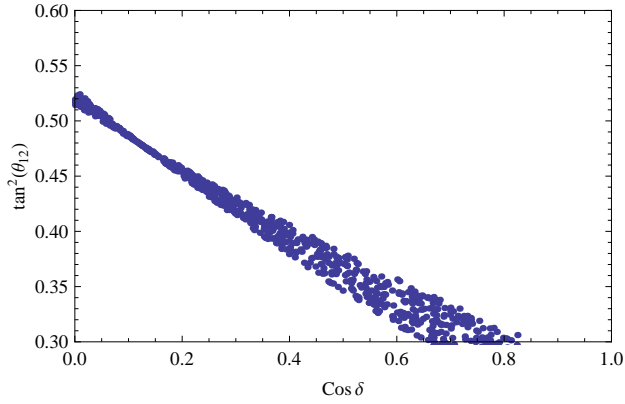}
\caption{Correlation plot between $\tan^2 \theta_{12}$ and $\cos \delta$ for
 TBM mixing under charged lepton correction in the $3\sigma$ range of $sin^{2}\theta_{13}$[6]}
\end{center}
\end{figure}

\begin{figure}
\begin{center}
\includegraphics[scale=1]{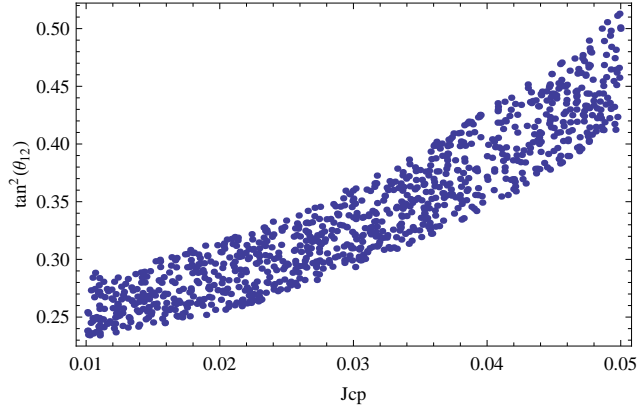}
\caption{Correlation plot between $\tan^2 \theta_{12}$ and $J_{CP}$ for TBM
  mixing under charged lepton correction in the range $0\leq \cos \delta \leq 1$}
\end{center}
\end{figure}

This PMNS matrix predicts 
\begin{equation}
\tan^2 \theta_{12}=\displaystyle \frac{\tilde{c}^2_{12}+\tilde{s}^2_{12}(\tilde{c}_{23}+\tilde{s}_{23})^2-2\tilde{c}_{12}\tilde{s}_{12}(\tilde{c}_{23}+\tilde{s}_{23})\cos \delta}
{2[\tilde{c}^2_{12}+\frac{1}{4}\tilde{s}^2_{12}(\tilde{c}_{23}+\tilde{s}_{23})^2+\tilde{c}_{12}\tilde{s}_{12}(\tilde{c}_{23}+\tilde{s}_{23})\cos \delta]}.
\end{equation}
The predictions on $\sin^2 \theta_{13}$ and $\tan^2 \theta_{23}$ remain unaffected by 
the phase $\delta$ and are given by eqs. (17) and (19) respectively. For
$\delta=0$, eq.(23) necessarily yields the same analytic expression given by
eq.(18). As discussed in subsection 2.1, the numerical value predicted by 
eq.(18) is much smaller than the global bestfit value but now the
dependency of $\tan^2 \theta_{12}$ on $\cos \delta$ can lift the prediction
on $\tan^2 \theta_{12}$ upto desired experimental 
prediction as shown in Fig.2. Fig.2 also shows that prediction on $\tan^2 \theta_{12}$ 
can accomodate $1\sigma$ and $3\sigma$ ranges of global data for non-zero
value of $\cos \delta$. For the best fit value $\tan^2 \theta_{12}=0.47$
we calculate $\cos \delta \approx 0.147$ from eq.(23).\\

The rephasing invariant quantity defined as $J_{CP}=Im\{U_{e2}U_{\mu 3}
U^{*}_{e3}U^{*}_{\mu 2}\}$, is obtained from the new PMNS matrix
$U^{\prime}_{PMNS}$ in eq.(22) as
\begin{equation}
|J_{CP}|=\displaystyle \frac{1}{6} \tilde{c}_{12} \tilde{s}_{12}(\tilde{c}_{23}
+\tilde{s}_{23})(\tilde{c}_{23}-\tilde{s}_{23})^2 \sin \delta.
\end{equation}

For maximal CP violation ($\delta=\frac{\pi}{2}$) and for numerical values
of $\tilde{s}_{12}$ and $\tilde{s}_{23}$ considered in subsection 2.1, eq.(24)
predicts $|J_{CP}|_{max}\approx 0.0359 $ while for $\cos \delta \approx 0.147$
we get $|J_{CP}|\approx 0.0355 $.\\ 

We would also like to analyze the structure of the PMNS matrix
under the criterion when a Dirac type CP phase $\phi$ is introduced
from the charged lepton sector. In this case the PMNS matrix can be
parametrized as 
\begin{equation}
U^{\prime \prime}_{PMNS}=(\tilde{R}_{23}.Diag(e^{i\phi},1,e^{-i\phi}).\tilde{R}_{12})^{\dagger}U_{TB},
\end{equation}
where $\tilde{R}_{12}$ and $\tilde{R}_{23}$ are given by eq.(14). Eq.(25)
then gives

\begin{equation}
U^{\prime \prime}_{PMNS}= \begin{pmatrix}
       \sqrt{\frac{2}{3}}[\tilde{c}_{12}e^{-i \phi}+ \frac{1}{2}\tilde{s}_{12}
                                 (\tilde{c}_{23}+\tilde{s}_{23})]  
     & \frac{1}{\sqrt{3}}[\tilde{c}_{12}e^{-i \phi}-\tilde{s}_{12}(\tilde{c}_{23}+\tilde{s}_{23})]  
           & -\frac{1}{\sqrt{2}}\tilde{s}_{12}(\tilde{c}_{23}-\tilde{s}_{23}) \\
   -\frac{1}{\sqrt{6}}[\tilde{c}_{12}(\tilde{c}_{23}+\tilde{s}_{23})-2\tilde{s}_{12}e^{-i \phi}] 
  & \frac{1}{\sqrt{3}}[\tilde{s}_{12}e^{-i \phi}+\tilde{c}_{12}(\tilde{c}_{23}+\tilde{s}_{23})]  
            & \frac{1}{\sqrt{2}}\tilde{c}_{12}(\tilde{c}_{23}-\tilde{s}_{23}) \\
              \frac{1}{\sqrt{6}}(\tilde{c}_{23}-\tilde{s}_{23})e^{i \phi}    
              &    -\frac{1}{\sqrt{3}}(\tilde{c}_{23}-\tilde{s}_{23})e^{i \phi}  
                  & \frac{1}{\sqrt{2}}(\tilde{c}_{23}+\tilde{s}_{23})e^{i \phi} \\

                \end{pmatrix}.
\end{equation}
This PMNS matrix leads to the following expressions for $\tan^2 \theta_{12}$
and $J_{CP}$:
\begin{equation}
\tan^2 \theta_{12}=\displaystyle \frac{\tilde{c}^2_{12}+\tilde{s}^2_{12}(\tilde{c}_{23}+\tilde{s}_{23})^2-2\tilde{c}_{12}\tilde{s}_{12}(\tilde{c}_{23}+\tilde{s}_{23})\cos \phi}
{2[\tilde{c}^2_{12}+\frac{1}{4}\tilde{s}^2_{12}(\tilde{c}_{23}+\tilde{s}_{23})^2+\tilde{c}_{12}\tilde{s}_{12}(\tilde{c}_{23}+\tilde{s}_{23})\cos \phi]},
\end{equation}
\begin{equation}
J_{CP}=\displaystyle \frac{1}{6} \tilde{c}_{12} \tilde{s}_{12}(\tilde{c}_{23}
+\tilde{s}_{23})(\tilde{c}_{23}-\tilde{s}_{23})^2 \sin \phi ,
\end{equation}
which are similar in structure to those given in eqs. (23) and (24). Further, from 
eqs. (17) and (20), eq.(24) can be approximated as 
\begin{equation}
|J_{CP}|\approx \displaystyle \frac{1}{3\sqrt{2}} \sin \theta_{13} \sin \delta
\end{equation}
which is consistent with the result of [9].

\section{Summary and Discussion}
We have discussed charged lepton correction to TBM mixing with a possible
form of $U_l$ which can generate $\sin \theta_{13}$ of the order of $\displaystyle
 \frac{\lambda}{\sqrt{2}}$ and $\tan^2 \theta_{23}<1$ under the consideration that
the charged lepton correction is CKM like.
The charged lepton mixing matrix $U_l$ is derived from three dimensional
rotation matrices in the same manner as the TBM neutrino mixing matrix.
We found that in the absence of CP violation numerical predictions on $\sin \theta_{13}$
and $\tan^2 \theta_{23}$ are consistent with latest global data but that on 
$\tan^2 \theta_{12}$ is significantly smaller than the global best fit value. However, when
we introduce the Dirac CP violating phase $\delta$, the expression for 
$\tan^2 \theta_{12}$ shows that it 
becomes dependent on  $\cos \delta$. This dependency can be employed to lift up
the value of $\tan^2 \theta_{12}$ to desired experimental prediction. 
For the best fit value $\tan^2 \theta_{12}=0.47$ we find 
$\cos \delta \approx 0.147$. Further we get expression for the rephasing
invariant quantity in case of TBM mixing which is consistent with the
result of [9].
 

\end{document}